\documentstyle[twoside,12pt]{article}

\catcode`\@=11
\long\def\@makefntext#1{ 
\protect\noindent \hbox to 3.2pt {\hskip-.9pt
$^{{\eightrm\@thefnmark}}$\hfil}#1\hfill} 

 \def\@makefnmark{\hbox to 0pt{$^{\@thefnmark}$\hss}}  

\def\ps@myheadings{\let\@mkboth\@gobbletwo
\def\@oddhead{\hbox{} 
\rightmark\hfil\eightrm\thepage}
\def\@oddfoot{}\def\@evenhead{\eightrm\thepage\hfil 
\leftmark\hbox{}}\def\@evenfoot{}
\def\sectionmark##1{}\def\subsectionmark##1{}}

\oddsidemargin=\evensidemargin
\newcounter{sectionc}\newcounter{subsectionc}\newcounter{subsubsectionc}
\renewcommand{\section}[1] {\vspace{12pt}\addtocounter{sectionc}{1}
\setcounter{subsectionc}{0}\setcounter{subsubsectionc}{0}\noindent
	{\bf\thesectionc. #1}\par\vspace{5pt}}
\renewcommand{\subsection}[1] {\vspace{12pt}\addtocounter{subsectionc}{1}
	\setcounter{subsubsectionc}{0}\noindent
	{\bf\thesectionc.\thesubsectionc. {\kern1pt \bf\it #1}}\par\vspace{5pt}}
\renewcommand{\subsubsection}[1] {\vspace{12pt}\addtocounter{subsubsectionc}{1}
	\noindent{\thesectionc.\thesubsectionc.\thesubsubsectionc.
	{\kern1pt \it #1}}\par\vspace{5pt}}
\newcommand{\nonumsection}[1] {\vspace{12pt}\noindent{\bf #1}
	\par\vspace{5pt}}

\newcommand{\textlineskip}{\baselineskip=14pt}
\newcommand{\smalllineskip}{\baselineskip=12pt}

\def\eightcirc{
\begin{picture}(0,0)
\put(4.4,1.8){\circle{6.5}}
\end{picture}}
\def\eightcopyright{\eightcirc\kern2.7pt\hbox{\eightrm c}}

\newcounter{itemlistc}
\newcounter{romanlistc}
\newcounter{alphlistc}
\newcounter{arabiclistc}

\newcommand{\fcaption}[1]{
        \addtocounter{figure}{1}
         {{\tenrm Fig.~\thefigure . #1} }\hfil\break }

\newcommand{\tcaption}[1]{			
        \addtocounter{table}{1}
         {{\tenrm\offinterlineskip Table~\thetable . #1} }\hfil\break }

\def\thebibliography#1{\nonumsection{\large \bf References}\list
  {[\arabic{enumi}]}{\settowidth\labelwidth{[#1]}\leftmargin\labelwidth
    \advance\leftmargin\labelsep
    \usecounter{enumi}}
    \def\newblock{\hskip .11em plus .33em minus .07em}
    \sloppy\clubpenalty4000\widowpenalty4000}

\def\pmb#1{\setbox0=\hbox{#1}
	\kern-.025em\copy0\kern-\wd0
	\kern.05em\copy0\kern-\wd0
	\kern-.025em\raise.0433em\box0}

\def\fnt#1#2{\footnotetext{\kern-.3em
	{$^{\mbox{\scriptsize #1}}$}{#2}}}

\def\fpage#1{\begingroup
\voffset=.3in
\thispagestyle{empty}\begin{table}[b]\centerline{\footnotesize #1}
	\end{table}\endgroup}

 1
 1
 1

\font\eightrm=cmr8

\def\qed{\hbox{${\vcenter{\vbox{                          
   \hrule height 0.4pt\hbox{\vrule width 0.4pt height 6pt
   \kern5pt\vrule width 0.4pt}\hrule height 0.4pt}}}$}}

\newcommand{\be}{\begin{eqnarray}}
\newcommand{\ee}{\end{eqnarray}}
\newcommand{\dslash}{\partial \hskip -0.5em /}
\newcommand{\Dslash}{D \hskip -0.7em /}

\newcommand{\tr}{{\rm tr}}
\newcommand{\Tr}{{\rm Tr}}

\newcommand{\A}{{\cal A}}

\newcommand{\T}{{\cal T}}
\newcommand{\ie}{{\it i.e.}\ }
\newcommand{\eg}{{\it e.g.}\ }

\input psfig

\begin{document}
\normalsize\textlineskip
{\thispagestyle{empty}
\setcounter{page}{1}

\fpage{1}
\rightline{UNITU-THEP-3/1993}
\rightline{April 1993}
\vspace{1cm}
\centerline{\Large \bf Strange and Non-Strange Meson Fluctuations}
\vspace{0.5cm}
\centerline{\Large \bf off the
Nambu--Jona-Lasinio Soliton$^\dagger $}
\vspace{0.9cm}
\centerline{H. Weigel, H. Reinhardt and R. Alkofer}
\vspace{0.9cm}
\centerline{Institute for Theoretical Physics}
\vspace{0.3cm}
\centerline{T\"ubingen University}
\vspace{0.3cm}
\centerline{Auf der Morgenstelle 14}
\vspace{0.3cm}
\centerline{D-7400 T\"ubingen, FR Germany}

\vspace{4cm}
\normalsize\textlineskip
\noindent
\centerline{\bf Abstract}
\vspace{0.5cm}

\noindent
Mesonic fluctuations off the chiral soliton of the Nambu--Jona-Lasinio
model are investigated. The hedgehog configuration is proven to
represent a local extremum of the action. The method is applied
to flavor SU(3) and the energy eigenvalue of the kaon bound state
in the soliton background is evaluated which is the key ingredient for
the Callan-Klebanov approach to hyperons. The energy eigenvalue of
the corresponding strange-valence-quark is found to be 183MeV higher
than the energy eigenvalue of the up-valence-quark when 400MeV
is assumed for the up-quark constituent mass.

\vfill

\noindent
$^\dagger $
{\footnotesize{Supported by the Deutsche Forschungsgemeinschaft (DFG) under
contract Re 856/2-1.}}
\eject

\normalsize\textlineskip
\leftline{\large \bf 1. Introduction}

\medskip

Chiral symmetry plays a central role in todays understanding of
light flavor hadrons. Especially, the properties of
pseudoscalar mesons, the would-be Goldstone bosons of
spontaneous chiral symmetry breaking can be described quite well
with the simplest chirally symmetric quark model, the
Nambu--Jona-Lasinio (NJL) model\cite{na61}.

In recent years it has been shown that the NJL model possesses
solitonic solutions\cite{re88a,me89,re88b,al90}. Applying the
well-known cranking procedure in flavor SU(2)\cite{ad83} to the
NJL model\cite{re89} allows for a description of the nucleon and the
$\Delta$ resonance\cite{go91,wa91}. It was
also demonstrated that the collective quantization within flavor SU(3)
and diagonalization of the resulting collective hamiltonian \`a la Yabu
and Ando\cite{ya88} reproduces the mass splitting of the $\frac{1}{2}^+$ and
$\frac{3}{2}^+$ baryons reasonably well\cite{we92b,we92a,bl92}.

In this letter we will develop a method for describing
mesonic fluctuations off the NJL soliton. First of all, this
demonstrates that the static hedgehog extremizes the action
(at least locally). This method is furthermore the basis
for describing certain properties of baryons, \eg $\pi-N$
scattering phase shifts\cite{sch89}, pion photo production,
nucleon polarizibilities\cite{br93} etc.. In this note we will consider
the kaon bound state of the soliton as an example since it
provides a useful tool to describe strange baryons. Such a
picture was extensively studied in Skyrme type models
\cite{ca85,ca88,bl89,ri91}.

\medskip

\leftline{\large \bf 2. Expansion of the Action}

\medskip

We assume the NJL action $\A_{NJL}$ with scalar and pseudoscalar
interactions as starting point which after bosonization\cite{eb86}
may be expressed as a sum $\A_{NJL}=\A_F+\A_m$ of a fermion determinant
\be
\A_F=\Tr\log(i\Dslash)=\Tr\log\left(i\dslash-(P_RM+P_LM^{\dag})\right)
\label{fdet}
\ee
and a purely mesonic part
\be
\A_m=\int d^4x\left(-\frac{1}{4G}\tr(M^{\dag}M-\hat m_0(M+M^{\dag}
)+\hat m_0^2)\right).
\label{ames}
\ee
wherein $P_{R,L}=(1\pm \gamma _5)/2$ denote the projectors on right- and
left-handed quark fields, respectively. The complex matrix $M=S+iP$
parametrizes the scalar and pseudoscalar meson fields,
$S_{ij}= S^a\lambda^a_{ij}/2$ and $P_{ij}= P^a\lambda^a_{ij} /2$.
We will consider the case of three flavors ($N_f=3$) and neglect
isospin breaking, \ie the current quark mass matrix acquires
the form $\hat m^0={\rm diag}(m^0,m^0,m^0_s)$ with
$m^0=(1/2)(m^0_u+m^0_d)$ being the average current quark mass
of up and down quarks. $G$ represents a coupling constant to be
determined later.

In order to regularize the diverging fermion determinant $\A_F=\A_R+\A_I$
this object is continued to euclidean space. An $O(4)$ invariant
cut-off $\Lambda$ is introduced by Schwinger's proper time
regularization\cite{sch51} which replaces the real part $\A_R$ by a parameter
integral
\be
\A_R=\frac{1}{2}\Tr\log\left(\Dslash_E^{\dag}\Dslash_E\right)
\longrightarrow -\frac{1}{2}\int_{1/\Lambda ^2}^\infty
\frac{ds}s\Tr\exp\left(-s\Dslash_E^{\dag}\Dslash_E\right) .
\label{arreg}
\ee
Note that (\ref{arreg}) becomes exact in the limit $\Lambda\rightarrow
\infty$ up to an irrelevant constant. Although the imaginary part
\be
\A_I=\frac{1}{2}\Tr\log\left((\Dslash_E^{\dag})^{-1}\Dslash_E\right)
\label{af3}
\ee
is finite we will later on also regularize $\A_I$ in a way consistent
with (\ref{arreg}).

In order to fix the parameters of the model we parametrize the
matrix $M$ in a way which will prove appropriate for discussing
fluctuations off the soliton:
\be
M=\xi_0\xi_f\Sigma\xi_f\xi_0.
\label{defm}
\ee
The matix $\Sigma$ is hermitian while $\xi_0$ and $\xi_f$ are unitary.
Note that this parametrization differs from the usually adopted
unitary gauge $M=\xi_L^\dagger\Sigma\xi_R,\ \xi_L^\dagger=\xi_R$.
In the baryon number zero sector $\xi_0$ is replaced by the unit matrix.
Varying the action with respect to scalar and pseudoscalar fields
yields the Schwinger-Dyson or gap equation which determines the vacuum
expectation value
$\langle \Sigma \rangle={\rm diag}(m,m,m_s)$
\be
m_i & = & m_i^0+m_i^3\frac{N_C G}{2\pi^2}
\Gamma\left(-1,(\frac{m_i}{\Lambda})^2\right)
\label{conmass}
\ee
wherein $m_i\ (i=u,d,s)$ denote the constituent quark masses and
$m=(m_u+m_d)/2$ the corresponding average value of up- and down
quarks. Space-time dependent fluctuating pseudoscalar meson
fields $\eta_a(x)$ are introduced via
\be
\xi_f(x)={\rm exp}\left(i\sum_{a=1}^8\eta_a(x)\lambda_a/2\right) .
\label{defeta}
\ee
The expression quadratic in $\eta_a$ provides formulae \cite{we92a}
for the pseudoscalar masses as well as the corresponding decay constants.
Using physical values $m_\pi=135{\rm MeV},\ m_k=495{\rm MeV}$ and
$f_\pi=93{\rm MeV}$
these expressions together with eqn.(\ref{conmass}) allow to fix
the parameters $\Lambda, G, m^0$ as well as the constituent mass of
the strange quark $m_s$ for a given value of $m$. The kaon decay
constant $f_k$ is left as a prediction. In table 1 we display the
numerical results only since these calculations have already been
reported and discussed in detail\cite{we92a}.

\begin{table}
\tcaption{Mass parameters fixed in the meson sector of the NJL model.
The kaon decay constant $f_k$ is predicted.}
\newline
\centerline{
\begin{tabular}{|c|c|c|c|}
\hline
$m$ (MeV) & $m_s$(MeV) & $m^0_s/m^0$ & $f_k$(MeV) \\
\hline
350&577&23.5&104.4\\
\hline
400&613&22.8&100.3\\
\hline
450&650&22.5& 97.4\\
\hline
500&687&22.3& 95.5\\
\hline
600&766&22.3& 93.1\\
\hline
800&934&22.8& 91.3\\
\hline
\end{tabular}}
\end{table}

For the ongoing calculations involving the static chiral soliton
$\xi_0(\mbox {\boldmath $r $})$ the meson fields will be constrained to
the chiral circle. We thus write for the matrix $M$
\be
M & = & \xi_0\xi_f\langle\Sigma\rangle\xi_f\xi_0.\nonumber \\*
\xi_0(\mbox {\boldmath $x $}) & = & {\rm exp}\left(\frac{i}{2}
{\mbox{\boldmath $\tau$}} \cdot{\bf \hat r}\ \Theta(r)\right).
\label{chsol}
\ee
The main goal of our calculations is to expand the action $\A$
in the presence of the soliton (\ref{chsol}) up to second order in the
space-time dependent meson fluctuations $\eta_a(x)$. To do so we
express the euclidean Dirac operator $\Dslash_E$ as
\be
i\beta\Dslash_E=-\partial_\tau-\mbox {\boldmath $\alpha \cdot p $}
-\T \beta\left(\xi_f\langle\Sigma\rangle\xi_f P_R
+\xi_f^\dagger\langle\Sigma\rangle\xi_f^\dagger P_L\right)\T^\dagger
\label{de1}
\ee
wherein $\tau=ix_0$ is the euclidean time. The unitary matrix
\be
\T=\xi_0 P_L + \xi_0^\dagger P_R
\label{deft}
\ee
contains the information about the chiral soliton. The appearance
of the meson fluctuations is most clearly exhibited by introducing
a hamiltonian $h$
in (\ref{de1})
\be
i\beta\Dslash_E=-\partial_\tau-h
=-\partial_\tau-\left(h_{(0)}+h_{(1)}+h_{(2)}+\cdot\cdot\cdot\right)
\label{de2}
\ee
wherein the subscript labels the power of the meson fluctuations:
\be
\!\!h_{(0)}({\bf r})&=&\!\!\mbox {\boldmath $\alpha \cdot p $}
-\T\beta\langle\Sigma\rangle\T^\dagger \nonumber \\*
\!\!&=&\!\!\mbox {\boldmath $\alpha \cdot p $} + \beta\left(
\frac{m}{\sqrt3}(\frac{2}{\sqrt3}+\lambda_8)({\rm cos}{\Theta}+
i{\mbox{\boldmath $\tau$}} \cdot{\bf \hat r}\gamma_5\ {\rm sin}\Theta)
+\frac{m_s}{\sqrt3}(\frac{1}{\sqrt3}-\lambda_8)\right)
\label{h0}
\\*
\!\!h_{(1)}({\bf r},t)&=&\!\!i\T\beta\gamma_5
\left(m \mbox {\boldmath $\eta \cdot \tau $}+\frac{1}{2}
(m+m_s)\sum_{\alpha=4}^7\eta_\alpha\lambda_\alpha\right)\T^\dagger
\label{h1}
\\*
\!\!h_{(2)}({\bf r},t)&=&\!\!-\T\beta\left(
\frac{m}{2\sqrt3}(\frac{2}{\sqrt3}+\lambda_8)
\mbox {\boldmath $\eta \cdot \eta $}+\frac{m+m_s}{8}
\Big\{\sum_{\alpha=4}^7\eta_\alpha\lambda_\alpha,
\sum_{\beta=4}^7\eta_\beta\lambda_\beta\Big\}\right)\T^\dagger .
\label{h2}
\ee
The meson field $\eta_8$ does not couple to the soliton and is hence
discarded. Obviously $h_{(i)}$ are hermitian operators. Noting
furthermore that $h_{(0)}$ is time independent we have up to
second order in the fluctuations
\be
\Dslash_E^{\dag}\Dslash_E=-\partial_\tau^2+h_{(0)}^2
-\big[\partial_\tau,h_{(1)}\big]+\big\{h_{(1)},h_{(0)}\big\}
-\big[\partial_\tau,h_{(2)}\big]+\big\{h_{(2)},h_{(0)}\big\}
+h_{(1)}^2+\cdot\cdot\cdot
\label{desq} .
\ee
It is useful to define a zeroth-order heat kernel operator\cite{re89,we92a}
$\hat K_0(s)={\rm exp}\left(s(\partial_\tau^2-h_{(0)}^2)\right)$
in order to expand $\A_R$
\be
\A_R &=& \A_R^{(0)}+\A_R^{(1)}+\A_R^{(2)}+\cdot\cdot\cdot
\label{arexpa} \\*
\A_R^{(0)} &=& -\frac{1}{2}\Tr\int_{1/\Lambda^2}^\infty
\frac{ds}{s}\hat K_0(s)
\label{ar0} \\*
\A_R^{(1)} &=& \frac{1}{2}\Tr\int_{1/\Lambda^2}^\infty ds
\hat K_0(s)\big\{h_{(1)},h_{(0)}\big\}
\label{ar1} \\*
\A_R^{(2)} &=& \frac{1}{2}\Tr\int_{1/\Lambda^2}^\infty ds
\hat K_0(s)\left(\big\{h_{(2)},h_{(0)}\big\}+h_{(1)}^2\right)
-\frac{1}{4}\Tr\int_{1/\Lambda^2}^\infty ds\int_0^s ds^\prime
\hat K_0(s-s^\prime)
\nonumber \\*
&&\qquad \times\left(\big[\partial_\tau,h_{(1)}\big]
\hat K_0(s^\prime)\big[\partial_\tau,h_{(1)}\big]+
\big\{h_{(1)},h_{(0)}\big\}\hat K_0(s^\prime)
\big\{h_{(1)},h_{(0)}\big\}\right).
\label{ar2}
\ee
Again extensive use has been made of
$\big[\partial_\tau,h_{(0)}\big]=0$. The cut-off $\Lambda$ in $\A_I$
is introduced by the substitution\cite{zu92}
\be
&&\left((\partial_\tau-h_{(0)}-h_{(1)})
(\partial_\tau+h_{(0)}+h_{(1)})\right)^{-1}
\nonumber \\*
&&\qquad\longrightarrow
-\int_{1/\Lambda^2}^\infty {\rm exp}\left(s(\partial_\tau-h_{(0)}-h_{(1)})
(\partial_\tau+h_{(0)}+h_{(1)})\right)
\ee
which is legitimate since the argument is negative definite, \ie it
converges for large momenta. The resulting imaginary part reads
\be
\A_I=\Tr\int_{1/\Lambda^2}^\infty ds\int_0^s ds^\prime
\hat K_0(s-s^\prime)\partial_\tau h_{(1)}
\hat K_0(s^\prime)h_{(0)}h_{(1)}+\cdot\cdot\cdot
\label{aiexpa}
\ee
which starts at second order in the fluctuations $\eta_a(x)$.

For carrying out the temporal part of the functional trace we perform
a Fourier transformation of the meson fluctuations in
euclidean space:
\be
\eta_a({\bf r},-i\tau)=\int_{-\infty}^{+\infty}
\frac{d\omega}{2\pi}\tilde \eta_a({\bf r},i\omega)
{\rm e}^{-i\omega\tau}
\label{feta}
\ee
which may be transferred to the hamiltonian:
\be
h_{(1)}({\bf r},-i\tau) & = & \int_{-\infty}^{+\infty}
\frac{d\omega}{2\pi}\tilde h_{(1)}({\bf r},i\omega)
{\rm e}^{-i\omega\tau}\quad {\rm and} \nonumber \\*
h_{(2)}({\bf r},-i\tau) & = & \int_{-\infty}^{+\infty}\frac{d\omega}{2\pi}
\int_{-\infty}^{+\infty}\frac{d\omega^\prime}{2\pi}
\tilde h_{(2)}({\bf r},i\omega,i\omega^\prime)
{\rm e}^{-i(\omega+\omega^\prime)\tau}
\label{fham}
\ee
wherein $\tilde h_{(i)}$ are obtained from $h_{(i)}$ (\ref{h1},\ref{h2})
through substitution of the meson fields by their Fourier transforms
(\ref{feta}). Note that for actual computations the frequency $\omega$
has to be continued back to Minkowski space. The temporal trace
essentially boils down to computing gaussian integrals. The spatial
part of the trace as well as the traces over Dirac and flavor indices
are evaluated using the eigenstates of the static one-particle
hamiltonian $h_{(0)}$
\be
h_{(0)} | \mu\rangle = \epsilon_\mu | \mu\rangle .
\label{diagham}
\ee
We should mention that this basis actually contains two independent
sets of eigenstates which are characterized by the strangeness
quantum number of the eigenstates. Those with zero strangeness
are functionals of the chiral angle $\Theta(r)$ while those with
strangeness $\pm1$ are free Dirac spinors.

The resulting formulae for the expansion of the fermion determinant
are found to be
\be
\A_R^{(0)}=-T\frac{N_C}{2}\int_{1/\Lambda^2}^\infty
\frac{ds}{\sqrt{4\pi s^3}}\sum_\mu
{\rm exp}(-s\epsilon_\mu^2)=-T E^{\rm vac}[\Theta].
\label{evac}
\ee
An infinitely large euclidean time interval $T$ is implicitly
assumed. $E^{{\rm vac}}[\Theta]$ is, obviously, just the sea
contribution of the fermion determinant to the soliton
mass\cite{re89}. Postponing the discussion of the linear term
\be
\A_R^{(1)}=N_C\int_{1/\Lambda^2}^\infty ds\ \sum_\mu
{\rm e}^{-s\epsilon_\mu^2}\epsilon_\mu
\langle\mu|\tilde h_{(1)}({\bf r},0)|\mu\rangle
\label{aflin}
\ee
to the subsequent section we may present the result for the quadratic
term $\A_F^{(2)}=\A_R^{(2)}+\A_I$ in Minkowski space
\be
\A_F^{(2)} & = & \frac{N_C}{2}\int_{1/\Lambda^2}^\infty
\frac{ds}{\sqrt{4\pi s}}\sum_\mu 2\epsilon_\mu
{\rm e}^{-s\epsilon_\mu^2}\int^{+\infty}_{-\infty}
\frac{d\omega}{2\pi}\langle\mu|\tilde h_{(2)}({\bf r},\omega,-\omega)
|\mu\rangle
\nonumber \\*
&&\!\!\!+\frac{N_C}{4}\int_{1/\Lambda^2}^\infty ds \sqrt{\frac{s}{4\pi}}
\sum_{\mu\nu}\int^{+\infty}_{-\infty}\frac{d\omega}{2\pi}
\langle\mu|\tilde h_{(1)}({\bf r},\omega)|\nu\rangle
\langle\nu|\tilde h_{(1)}({\bf r},-\omega)|\mu\rangle
\label{afquad}
\\*
&&\times \left\{\frac{{\rm e}^{-s\epsilon_\mu^2}
+{\rm e}^{-s\epsilon_\nu^2}}{s}
+[\omega^2-(\epsilon_\mu+\epsilon_\nu)^2]
R_0(s;\omega,\epsilon_\mu,\epsilon_\nu)
-\omega\epsilon_\nu R_1(s;\omega,\epsilon_\mu,\epsilon_\nu)
\right\}.
\nonumber
\ee
The last term, containing only odd powers of $\omega$, stems from
the imaginary part. The regulator functions $R_i$ involve
Feynman parameter integrals reflecting the quark loop in
the presence of the soliton
\be
R_i(s;\omega,\epsilon_\mu,\epsilon_\nu)
=\int_0^1 x^i dx\ {\rm exp}\left(-s[(1-x)\epsilon_\mu^2
+x\epsilon_\nu^2-x(1-x)\omega^2]\right)\quad (i=0,1).
\label{regfct}
\ee

Besides the Dirac sea also the explicit occupation of the valence
quark level contributes to the action as long as the corresponding
energy eigenvalue $\epsilon_{\rm val}$ is positive. Since no
regularization is involved the computation is completely performed
in Minkowski space. Treating the meson fluctuations as time-dependent
perturbations the associated first order change $\delta \Psi_{\rm val}$
of the valence quark wave-function $\Psi_{\rm val}$ is obtained to be
\be
\delta \Psi_{\rm val}({\bf r},t)=
\left(i\partial_t-h_{(0)}({\bf r})\right)^{-1}
h_{(1)}({\bf r},t)\Psi_{\rm val}({\bf r},t).
\label{delpsi}
\ee
The corresponding contribution to the mesonic action reads
\be
\A_{\rm val} & = & -\eta^{\rm val}N_C\Bigg\{T \epsilon_{\rm val}
+\langle{\rm val}|\tilde h_{(1)}({\bf r},0)|{\rm val}\rangle
+\int^{+\infty}_{-\infty}\frac{d\omega}{2\pi}
\Big(\langle{\rm val}|\tilde h_{(2)}({\bf r},\omega,-\omega)
|{\rm val}\rangle \nonumber \\*
&&\qquad\qquad\qquad
+\sum_{\mu\ne{\rm val}}\frac{
\langle{\rm val}|\tilde h_{(1)}({\bf r},\omega)|\mu\rangle
\langle\mu|\tilde h_{(1)}({\bf r},-\omega)|{\rm val}\rangle}
{\epsilon_{\rm val}-\omega-\epsilon_\mu}\Big)\Bigg\} .
\label{valquad}
\ee
Here $\eta^{\rm val}=0,1$ denotes the occupation number of
the valence quark and anti-quark states. The valence quark
also contributes terms of odd power in $\omega$ to the
action. As for the contribution of the Dirac sea (\ref{afquad})
these terms correspond to the imaginary part in euclidean space.
Terms of odd power in $\omega$ have the important property of
removing the degeneracy between solutions with $\pm\omega$.

Finally we need to expand the purely mesonic part of the
action (\ref{ames}). This proceeds analogously to Skyrme
model calculations resulting in
\be
\A_m & = & -T m_\pi^2f_\pi^2\int d^3r\
\left(1-{\rm cos}\Theta\right)
-m_\pi^2f_\pi^2\int d^3r\ {\rm sin}\Theta\
\hat {\bf r}\cdot\tilde{\mbox{\boldmath $\eta$}}(0)
\nonumber \\*
&&-\frac{1}{2}m_\pi^2f_\pi^2\int d^3r
\int^{+\infty}_{-\infty}\frac{d\omega}{2\pi}
\Big\{{\rm cos}\Theta\ \tilde{\mbox{\boldmath $\eta$}}(\omega)
\cdot\tilde{\mbox{\boldmath $\eta$}}(-\omega)
\nonumber \\*
&&\qquad\qquad\qquad
+\frac{1}{4}\Big(1+\frac{m_s}{m}\Big)
\Big({\rm cos}\Theta\ +\frac{m_s^0}{m^0}\Big)
\sum_{\alpha=4}^7\tilde\eta_\alpha(\omega)
\tilde\eta_\alpha(-\omega)\Big\}
\label{amquad}
\ee
wherein we made use of the relation
$G= m^0 M/m_\pi^2f_\pi^2$\cite{eb86}.

\medskip

\leftline{\large \bf 3. The linear Term}

\medskip

The zeroth-order hamiltonian $h_{(0)}$ commutes with the grand spin
operator $\mbox{\boldmath $G$}=\mbox{\boldmath $l$}+
\mbox{\boldmath $\sigma$}/2+\mbox{\boldmath $\tau$}/2$. Thus the
eigenstates $|\mu\rangle$ in (\ref{diagham}) are degenerate with respect
to the grand spin projection quantum number $M_G$. The sum over
$M_G$ in (\ref{aflin}) and in the linear term of eqn. (\ref{valquad})
projects out the grand spin zero piece of $\tilde h_{(1)}$
\be
\hat P_{G=0}\Big(\tilde h_{(1)}({\bf r},0)\Big)=
\hat M \beta\left(-{\rm sin}\Theta+
i{\mbox{\boldmath $\tau$}} \cdot\hat{\bf r}\
\gamma_5{\rm cos}\Theta\right) \hat P_{L=0}\Big(\hat{\bf r}
\cdot\tilde{\mbox{\boldmath $\eta$}}({\bf r},0)\Big).
\label{pg0h1}
\ee
$\hat P_{G=0}\Big(\tilde h_{(1)}({\bf r},0)\Big)$ is even under
parity transformations since $\hat P_{L=0}\Big(\hat{\bf r}
\cdot\tilde{\mbox{\boldmath $\eta$}}({\bf r},0)\Big)$ reduces to a
radial function. Noting futhermore that
$\tilde{\mbox{\boldmath $\eta$}}({\bf r},0)
=\int dt\ {\mbox{\boldmath $\eta$}}({\bf r},t)$,
the linear piece of the action collects up to
\be
\A^{(1)} & = &N_C m \int d^4x\ \hat P_{L=0}\Big(\hat{\bf r}\cdot
{\mbox{\boldmath $\eta$}}({\bf r},t)\Big)
\nonumber \\*
&&\times\Big\{
{\rm sin}\Theta\Big[-\frac{m_\pi^2f_\pi^2}{N_C m}
+\tr\Big(\sum_\mu {\rm sign}(\epsilon_\mu){\cal N}_\mu
\Psi_\mu({\bf r})\bar\Psi_\mu({\bf r})
+\eta^{\rm val}\Psi_{\rm val}({\bf r})\bar\Psi_{\rm val}({\bf r})\Big)\Big]
\nonumber \\*
&&-{\rm cos}\Theta\
\tr\ \Big[i{\mbox{\boldmath $\tau$}} \cdot\hat{\bf r}\ \gamma_5
\Big(\sum_\mu {\rm sign}(\epsilon_\mu){\cal N}_\mu
\Psi_\mu({\bf r})\bar\Psi_\mu({\bf r})+\eta^{\rm val}
\Psi_{\rm val}({\bf r})\bar\Psi_{\rm val}({\bf r})\Big)\Big]\Big\}.
\label{atotlin}
\ee
Here the trace runs over Dirac and isospin indices only.
${\cal N}_\mu =-\frac{1}{\sqrt\pi}\Gamma\big(\frac{1}{2},
(\epsilon_\mu/\Lambda)^2\big)$ refers to the  vacuum ``occupation
numbers" in the proper time regularization scheme\cite{re88b}.
The expression in curely brackets in eqn. (\ref{atotlin}) may easily
be verified to be the equation of motion for the chiral
angle\cite{re88a,we92a}. Thus the linear term vanishes for the
self-consistent chiral soliton. This proves that the hedgehog,
at least, represents a local extremum of the NJL action. The
main ingredient of this prove is the completeness of the basis
(\ref{diagham}), \ie the fact that the functional trace may be
computed using the eigenstates of $h_{(0)}$.

\medskip

\leftline{\large \bf 4. Kaon Fluctuations}

\medskip

Next we wish to apply the approach developed in section 2 to
the well known bound state approach. {\it I.e.} in the presence of
SU(3) symmetry breaking we wish to evaluate the energy eigenvalue
of the kaon bound by the soliton. This treatment was
originally set forth by Callan and Klebanov for the Skyrme model
to describe strange baryons\cite{ca85}. Thus we consider
meson fluctuations into strange direction only
\be
\mbox{\boldmath $\eta$}({\bf r},t)=0 \quad {\rm and} \quad
\sum_{\alpha=4}^7\eta_\alpha({\bf r},t)\lambda_\alpha=
\pmatrix{0 & K({\bf r},t)\cr K^\dagger({\bf r},t)&0\cr}
\label{stfluc}
\ee
wherein $K({\bf r},t)$ is a two-component iso-spinor. The
corresponding Fourier transform reads
\be
\sum_{\alpha=4}^7\tilde\eta_\alpha({\bf r},\omega)\lambda_\alpha=
\pmatrix{0 & \tilde K({\bf r},\omega)\cr
\tilde K^\dagger({\bf r},-\omega)&0\cr}.
\label{stft}
\ee

In the case of SU(3) symmetry ($m_s^0=m^0$) the associated
zero-mode of the fluctuation matrix (see below) corresponds to
\be
K_0({\bf r})={\bf \hat r}\cdot{\mbox{\boldmath $\tau$}} U_0
\pmatrix{{\rm sin}\frac{\Theta(r)}{2} \cr 0\cr}
\label{zeromode}
\ee
where $U_0$ is an arbitrary $2\times2$ space-time independent unitary
matrix fixing the isospin orientation. Eqn. (\ref{zeromode}) is obtained
by parametrizing the infinitesimal vector transformation of the chiral
soliton (\ref{chsol}) into strange direction in terms of the
fluctuation (\ref{stfluc}). $K_0({\bf r})$ obviously carries
isospin $T=1/2$, orbital angular momentum $L=1$ and grand spin
$G=1/2$. A kaon bound state develops in this channel once SU(3) symmetry
is abandoned and physical parameters ({\it cf.} table 1) are assumed.
We thus employ the ansatz
\be
\tilde K({\bf r},\omega)={\bf \hat r}\cdot{\mbox{\boldmath $\tau$}}
\Omega(r,\omega)
\label{bound}
\ee
for the kaon bound state. $\Omega(r,\omega)$ is
a two-component iso-spinor which only depends on  the radial
coordinate $r$ and the frequency $\omega$, \ie the angular
dependence is separated. Using eqn. (\ref{bound}) the
perturbative parts of the hamiltonian acquire a simple shape
\be
\!\!\!\tilde h_{(1)}({\bf r},\omega)&=&-\frac{1}{2}
\big(m+m_s\big)\beta \nonumber \\*
&&\times
\pmatrix{0 & {\hspace{-2cm}}\left({\rm sin}\frac{\Theta}{2}
-i{\bf \hat r}\cdot{\mbox{\boldmath $\tau$}}\gamma_5
{\rm cos}\frac{\Theta}{2}\right)
\Omega(r,\omega)\cr
\Omega^\dagger(r,-\omega)\left({\rm sin}\frac{\Theta}{2}-
i{\bf \hat r}\cdot{\mbox{\boldmath $\tau$}}\gamma_5
{\rm cos}\frac{\Theta}{2}\right)&0\cr}
\label{h1str}
\\*
\!\!\tilde h_{(2)}({\bf r},\omega,-\omega)&=&
\frac{1}{4}\big(m+m_s\big)^2\beta \nonumber \\*
&&{\hspace{-3cm}}\times\!
\pmatrix{\left({\rm sin}\frac{\Theta}{2}-i{\bf \hat r}\cdot
{\mbox{\boldmath $\tau$}}\gamma_5{\rm cos}\frac{\Theta}{2}\right)
\Omega(r,\omega)\Omega^\dagger(r,\omega)
\left({\rm sin}\frac{\Theta}{2}-i{\bf \hat r}\cdot
{\mbox{\boldmath $\tau$}}\gamma_5
{\rm cos}\frac{\Theta}{2}\right)&0\cr
0&{\hspace{-2cm}}-\Omega^\dagger(r,-\omega)\Omega(r,-\omega)\cr}
\label{h2str}
\ee

The main task now consists of computing matrix elements of the from
\be
\big|\ {_{\rm n s}\langle}\mu|\tilde h_{(1)}|\nu\rangle_{\rm s}\ \big|^2,
\qquad{_{\rm n s}\langle}\mu|\tilde h_{(2)}|\nu\rangle_{\rm n s}\
\quad {\rm and}\quad
{_{\rm s}\langle}\mu|\tilde h_{(2)}|\nu\rangle_{\rm s}
\label{matel}
\ee
with  $|\mu\rangle_{\rm n s}=\sum_\rho V_{\rho\mu}|\rho\rangle_{\rm n s}^0$
denoting SU(2) hedgehog states while
$|\nu\rangle_{\rm s}=|\nu\rangle_{\rm s}^0$ have non-vanishing strangeness.
The superscript refers to the free basis described in appendix B of
ref.\cite{we92a}. The orthogonal matrix $V$ is obtained by
diagonalizing $h_{(0)}$ in the non-strange sector and solving the
equation of motion for the hedgehog ({\it cf.} eqn. (\ref{atotlin}))
self-consistently\cite{re88a,me89,re88b,al90,we92a}. The evaluation of
the matrix elements (\ref{matel}) is as straightforward as painful,
however, the resulting expression for the action may generically written as
\be
\A[\Omega]&=&
\int_{-\infty}^{+\infty}\frac{d\omega}{2\pi}\Big\{\int drr^2
\int dr^\prime r^{\prime2}\ \Phi^{(2)}(\omega;r,r^\prime)
\Omega^\dagger(r,\omega)\Omega(r^\prime,\omega)
\nonumber \\*
&&\qquad\qquad\qquad
+\int dr r^2\ \Phi^{(1)}(\omega;r)\Omega^\dagger(r,\omega)
\Omega(r,\omega)\Big\}
\label{kernel}
\ee
The bilocal kernel $\Phi^{(2)}(\omega;r,r^\prime)$ stemming from the terms
quadratic in $h_{(1)}$ is symmetric under the exchange of $r$ and
$r^\prime$. The terms involving $h_{(2)}$ as well as the mesonic part
of the action (\ref{amquad}) contribute to the local kernel
$\Phi^{(1)}(\omega;r)$. The fact that these two kernels represent
unit matrices in iso-space simply reflects isospin invariance.
The equation of motion for $\Omega(r,\omega)$ then turns out to be
a homogeneous linear integral equation
\be
r^2\left\{\int dr^\prime r^{\prime2} \Phi^{(2)}(\omega;r,r^\prime)
\Omega(r^\prime,\omega)+\Phi^{(1)}(\omega;r)\Omega(r,\omega)
\right\}=0
\label{eqmfluc}
\ee
which in fact is a Bethe-Salpeter equation in a soliton background
and corresponds to the bound state equation of the Callan-Klebanov
approach\cite{ca85,ca88} to the Skyrme model.

$K({\bf r},t)$ represents an $s\bar q\ (\bar s q)$ excitation which,
according to eqn. (\ref{delpsi}), converts a non-strange valence
quark into a strange valence (anti-) quark. The emerging strange
valence (anti-) quark carries total angular momentum $J=1/2$ and
isospin $I=0$. The spinor representation reads
\be
\delta \Psi_{\rm val}^s({\bf r},t)
&=&\int_{-\infty}^{+\infty}\frac{d\omega}{2\pi}
\delta\tilde\Psi_{\rm val}^s({\bf r},\omega)\
{\rm e}^{-i(\epsilon_{\rm val}-\omega)t}
\label{psift} \\*
\delta\tilde\Psi_{\rm val}^s({\bf r},\omega)&=&
-\frac{1}{2}(\hat M+M_s)
\left(\epsilon_{\rm val}-\omega-h_{(0)}({\bf r})\right)^{-1}
\nonumber \\*
&&\qquad\qquad
\times\Omega^\dagger(r,\omega)
\left({\rm sin}\frac{\Theta}{2}-
i{\bf \hat r}\cdot{\mbox{\boldmath $\tau$}}\gamma_5
{\rm cos}\frac{\Theta}{2}\right)\Psi_{\rm val}^{\rm ns}({\bf r})
\label{psistr}
\ee
wherein we have separated the time dependence of the non-strange
valence quark. The interpretation of eqns. (\ref{psift}) and
(\ref{psistr}) is obvious: A kaon fluctuation with frequency $\omega$
changes a non-strange valence quark of energy $\epsilon_{\rm val}$
into a strange valence (anti-) quark with the energy
$\epsilon_{\rm val}^{\rm s}=\epsilon_{\rm val}-\omega$.
For $\epsilon_{\rm val}^{\rm s}>0$ this state has to be identified
with a strange quark carrying strangeness $S=-1$ while for
$\epsilon_{\rm val}^{\rm s}<0$ an $S=+1$ anti-quark is induced.
For $\epsilon_{\rm val}>0$ a baryon with vanishing strangeness
is constructed by occupying the valence quark state $N_C$-times.
The contribution of the valence quarks to the energy of this ($S=0$)
baryon is therefore $N_C \epsilon_{\rm val}$. The occupation of a
kaon mode with frequency $\omega$ replaces this non-strange
baryon by a baryon with strangeness -1 (+1) for
$\epsilon_{\rm val}^{\rm s}>0$ ($\epsilon_{\rm val}^{\rm s}<0$).
The corresponding contribution of the valence quarks to the energy
of this hyperon hence is $N_C \epsilon_{\rm val}-\omega$.
This discussion reveals that a ${\underline{\rm negative}}$
bound state energy $\omega$ is required in order to describe
physical hyperons.

Eqn. (\ref{psistr}) also suggests a suitable normalization of
the kaon field
\be
\int d^3r\
\left(\delta\tilde\Psi_{\rm val}^s({\bf r},\omega)\right)^\dagger
\delta\tilde\Psi_{\rm val}^s({\bf r},\omega) = 1
\label{norm}
\ee
since the overall factor for $\Omega(r,\omega)$ cannot be fixed
by the bound state equation (\ref{eqmfluc}).

\medskip

\leftline{\large \bf 5. Numerical results}

\medskip

The first step to solve the integral equation for the kaon fluctuation
numerically consists of discretizing eqn. (\ref{eqmfluc}):
$r_i=i\Delta r$ with $i=0,...,N$. Next, this equation is extended
to an eigenvalue problem
\be
r_i^2\sum_{j>0}\left\{\Delta r\ r_j^2\ \Phi^{(2)}_{ij}(\omega)
+\delta_{ij}\Phi^{(1)}_{j}(\omega)\right\}\Omega_j(\omega)
=\lambda\Omega_i(\omega)
\label{eigprob}
\ee
for $i>0$. The coefficient matrix on the $LHS$ is symmetric possessing
real eigenvalues $\lambda$ which are computed for a given frequency
$\omega$. Finally $\omega$ is tuned such that (\ref{eigprob})
allows for a solution with the eigenvalue $\lambda=0$. For this case also
\be
\Omega_0(\omega)=-\frac{1}{\Phi^{(1)}_{0}(\omega)}
\sum_{j>0}\Delta r\ r_j^2\ \Phi^{(2)}_{0j}(\omega)\Omega_j(\omega)
\label{omega0}
\ee
may be evaluated since neither $\Phi^{(2)}(\omega;r,r^\prime)$ nor
$\Phi^{(1)}(\omega;r)$ vanish at the origin. The resulting eigenvector
$\Omega_i(\omega)$ is identified as the discretized solution of eqn.
(\ref{eqmfluc}) and the corresponding frequency $\omega$ is the bound
state energy we are after. Fortunately the existence of a zero-mode in
the SU(3) symmetric case provides us with an excellent possibility to
test the precision of our numerical results. In figure 1 we compare
the radial dependence of our result (\ref{eigprob},\ref{omega0}) in the
SU(3) symmetric case with the zero-mode radial function ${\rm sin}\Theta/2$.
For the parameters of the numerical calculation we have chosen $N=17$
and $r_N=2.5fm$.
\begin{figure}
\centerline{\hskip -1.5cm
\psfig{figure=bzero.tex.ps,height=9.0cm,width=16.0cm}}
\fcaption{The radial dependence of the bound state wave-function
for various strength of the symmetry breaking for the case
of a constituent quark mass $m=400$MeV. Also shown is
the exact zero-mode. The normalization is chosen such that all
wave-functions coincide at $r=0$.}
\end{figure}
{}From figure 1 we observe a good agreement between the numerical and
analytical result for the zero mode (The numerical zero-mode
corresponds to $m_k=m_\pi=135$MeV.). However, we should remark that
in our calculation for the SU(3) symmetric case the numerical value
for the frequency $\omega$ is not exactly vanishing but one
has $\omega=-1.4$MeV. This shows that the inaccuarcy of the
numerical results is more than two orders of magnitude lower
than the inherited mass scale $m=400$MeV and thus negligible.
Figure 1 also demonstrates how the bound state evolves and gets more
strongly located at the origin as the symmetry breaking increases.

In table 2 we finally present our results for the bound state
energy $\omega$.
\begin{table}[b]
\tcaption{The valence quark energy of the non-strange
($\epsilon_{\rm val}$) and strange quarks $\epsilon_{\rm val}^{\rm s}$
as well as the bound state energy $\omega$ as functions of the
non-strange constituent quark mass $m$. Also shown is the
strange constituent quark mass $m_s$. Our input parameters are
those displayed in table 1. All numbers are in MeV.}
\newline
\centerline{\tenrm\smalllineskip
\begin{tabular}{|l|c c c c c|}
\hline
$m$ & 350 & 400 & 450 & 600 & 800 \\
\hline
$m_s$ & 577 & 613 & 650 & 766 & 934  \\
\hline
$\epsilon_{\rm val}$ & 249 & 211 & 181 & 99 & -19 \\
\hline
$\omega$   & -207& -183& -164 & -126 & -58 \\
\hline
$\epsilon_{\rm val}^{\rm s}$ & 456 & 394 & 345 & 225 & 37 \\
\hline
\end{tabular}}
\end{table}
In consideration of the discussion proceeding eqn. (\ref{psistr})
we find that the kaon bound state in NJL model describes
real hyperons carrying strangeness $S=-1$. And even more: For the
value obtained using physically reasonable values of the
non-strange constituent quark mass $m\approx400$MeV
each occupation of the kaon bound state increases the energy
of the hyperon by about 180MeV. Noting that the mass difference
between the nucleon and $\Lambda$ hyperon is just 177MeV we
observe that our result is just in the right ballpark. However,
we should remark that we have not yet carried out the quantization
of spin and isospin which usually is performed by introducing
collective coordinates for the corresponding zero modes\cite{ad83,ca85}.
This problem is subject to future investigations. We also would like to
remark that we do not find a solution in the interval $0<\omega<m_k$
for  $m < 600$MeV while for $m=700$MeV an additional bound state at
$\omega\approx350$MeV appears. Such a bound state describes exotic
baryons with $S=+1$ not available in the quark model. However, as may
be deduced form table 1, such large values for $m$ are not suitable
to describe physical systems since already in the meson sector we
find that \eg the kaon decay constant is incorrectly predicted to be
of the order of or even smaller than the pion decay constant.

\begin{figure}
\centerline{\hskip -1.5cm
\psfig{figure=bstate.tex.ps,height=9.0cm,width=16.0cm}}
\fcaption{The radial part of the kaon bound state for
several values of the constituent quark mass $m$.}
\end{figure}
In figure 2 we display the radial part of bound state wave-function
$\Omega(r,\omega)$ for several constituent quark masses. The
wave-function is normalized according to eqn. (\ref{norm}).
As expected, a significant localization shows up. This localization
may also be recognized from the induced strange quark wave-function.
The corresponding upper and lower components are presented in
figures 3 and 4, respectively. For the strange quark
wave-function $\Psi^{\rm s}_{\rm val}$ we also observe the
expected behavior that the lower component is the stronger suppressed
the larger the energy $\epsilon_{\rm val}^{\rm s}$ turns out to be.
\begin{figure}
\centerline{\hskip -1.5cm
\psfig{figure=upper.tex.ps,height=9.0cm,width=16.0cm}}
\fcaption{The radial part of the upper component of the
induced strange valence quark for several values of the
constituent quark mass $m$.}
\end{figure}

\begin{figure}
\centerline{\hskip -1.5cm
\psfig{figure=lower.tex.ps,height=9.0cm,width=16.0cm}}
\fcaption{The radial part of the lower component of the
induced strange valence quark for several values of the
constituent quark mass $m$.}
\end{figure}

\medskip

\leftline{\large \bf 6. Summary and Outlook}

\medskip

We have developed a method to describe meson fluctuations off the
chiral soliton of the NJL model. The vanishing linear term shows that
the hedgehog is (at least) a local extremum of the action. Furthermore
in the generalization to flavor SU(3) the result for the kaon bound state
energy indicates that applying the Callan-Klebanov approach\cite{ca85}
to the NJL model may provide a suitable treatment to describe hyperons.
We should mention that, in contrast to Skyrme model calculations\cite{ca88},
a negative bound energy is needed to describe physical hyperons. The
reason for this difference is the existence of explicit strange valence
quarks(\ref{psift})}.

We have not yet investigated the $1/N_C$ corrections which allow to
project onto physical baryons. In order to do so collective coordinates
for the (iso-) rotational zero-modes have to be introduced. Besides the
corresponding moment of inertia\cite{re89,go91,wa91,we92a} a coupling
between the bound state and the collective coordinates will show
up\cite{ca85}. The latter removes the degeneracy between baryons
with identical spin but different isospin, as \eg $\Lambda$ and
$\Sigma$. For the calculation in the NJL model an expansion of the
fermion determinant up to third order in the perturbation of the
hamiltonian has to be performed. We have started on this extensive
computation.

Furthermore a variety of applications for the treatment proposed in
this letter exists. {\it E.g.} in SU(2) one might \eg want to evaluate
the corrections to the soliton mass due to quantum fluctuations
analogously to Skyrme model calculations\cite{mo91}. The treatment is
also relevant for the computation of response functions appearing in the
exploration of nucleon polarizibilities\cite{br93}. The investigation
of the pionic fluctuation may be used for the calculation of the
$\pi-N$ scattering phase shifts\cite{sch89}.

\end{document}